\documentclass{aa}
\usepackage{graphicx,natbib}
\bibpunct{(}{)}{;}{a}{}{,}
\usepackage{txfonts}
\begin{document}

\title{Dust distribution in protoplanetary disks}
\subtitle{Vertical settling and radial migration}
\author{L. Barri\`ere-Fouchet\inst{1} \and J.-F. Gonzalez\inst{1}
\and J. R. Murray\inst{2} \and R. J. Humble\inst{3} \and S. T. Maddison\inst{2}}
\institute{Centre de Recherche Astronomique de Lyon (CNRS-UMR 5574),
\'Ecole Normale Sup\'erieure de Lyon, 46 all\'ee d'Italie,
\mbox{F-69364} Lyon C\'edex 07, France
\\e-mail: {\tt Laure.Barriere@ens-lyon.fr, Jean-Francois.Gonzalez@ens-lyon.fr}
\and
Centre for Astrophysics and Supercomputing, Swinburne University
of Technology, PO Box 218, Hawthorn, VIC~3122, Australia\\
e-mail: {\tt jmurray@swin.edu.au, smaddison@swin.edu.au}
\and
Canadian Institute for Theoretical Astrophysics, University of Toronto,
60, St. George Street, Toronto, ON M5S 3H8, Canada\\
e-mail: {\tt rjh@cita.utoronto.ca}}
\offprints{L. Barri\`ere-Fouchet}

\date{Received 25 October 2004 / Accepted 3 August 2005}

\abstract{We present the results of a three dimensional, locally isothermal,
non-self-gravitating SPH code which models protoplanetary disks with two
fluids: gas and dust. We ran simulations of a 1~$M_\odot$ star surrounded
by a 0.01~$M_\odot$ disk comprising 99\% gas and 1\% dust in mass and extending from
0.5 to $\sim 300$~AU. The grain size ranges from $10^{-6}$~m to $10$~m for the
low resolution ($\sim 25\,000$ SPH particles) simulations and from $10^{-4}$~m
to $10$~cm for the high resolution ($\sim 160\,000$ SPH particles) simulations.
Dust grains are slowed down by the sub-Keplerian gas and lose angular
momentum, forcing them to migrate towards the central star and settle to
the midplane. The gas drag efficiency varies according to the grain size,
with the larger bodies being weakly influenced and following marginally
perturbed Keplerian orbits, while smaller grains are strongly coupled
to the gas. For intermediate sized grains, the drag
force decouples the dust and gas, allowing the dust to preferentially
migrate radially and efficiently settle to the midplane.
The resulting dust distributions for each grain size will indicate, when
grain growth is added, the regions when planets are likely to form.
}

\maketitle

\section{Introduction}

Protoplanetary disks are by definition where we expect planets to
form, with micron to millimetre size grains characteristic of the
interstellar medium collecting and aggregating to form bodies
thousands of kilometres in diameter. Grain growth initially occurs via
individual grains sticking via collisions. However, once grains
reach millimetre size, their collision velocities are sufficiently large
to shatter the grains upon impact \citep{1996ApJ...469..740J}.
One way of reducing the relative velocity of colliding grains is to increase
their number density. Until recently, the dust dynamics of protoplanetary
disks has mostly been studied from the viewpoint of calculating rates of
radial migration into the central star. Classical work in this field was done
by \citet{1977MNRAS.180...57W}, who investigated the nature of the drag force
between the dust and gas components of a non-turbulent protostellar disk
and then derived equations describing the radial motions of individual
dust particles. Weidenschilling concluded that the maximum
radial velocity was independent of the drag law and was also insensitive
to the nebula mass. Radial migration was fastest for meter sized
objects with a velocity of $\sim 10^4$~cm\,s$^{-1}$.

\citet{1996A&A...309..301S,1997A&A...319.1007S} added turbulence
and grain coagulation and stressed the idea that the dust
motion is decoupled from the gas for a given range of particle size.
They found that grains with sizes 0.1 to $10^4$~cm have inward
radial velocities larger than that of the gas, while larger particles
have smaller velocities. \citet{1997A&A...319.1007S} modelled a Minimum 
Mass Solar Nebula (0.24~$M_\odot$) with radial extent 15~AU in which
all the solids migrated onto the star, whereas a nebula of 0.002~$M_\odot$ and
extending over 250~AU (which is close to our disk parameters) resulted in 
a distribution of solids close to that of our solar system. They did
not investigate vertical settling to the midplane.

\citet{Rice2004} studied the concentration of dust in the spiral arms of
marginally stable, self-gravitating protoplanetary disks. They followed the
evolution of dust test particles added into their existing SPH code which
models gas accretion disks. The test particles feel the gravitational effects
of the star and gas disk, as well as gas drag, but do not influence the gas
disk. They found that the dust density enhancement was significant only for
particle sizes between 10 and 100~cm with their nebula parameters, suggesting
that grain growth was possible due to the increased density in the dust layer
and the decrease in the relative velocities of the dust grains to each other.
Vertical settling was found to be insignificant inside the spiral arms.

The case of vertical settling was investigated by
\citet{2004ApJ...603..292G}, who analytically derive fluid equations for the 
average motion of particles in a non-turbulent nebula. 
Starting from the momentum equation of a single particle in
either Stokes or Epstein drag regime (for respectively large and small
particles), a Boltzmann distribution is used to obtain the
collective behaviour of dust particles which is consistent as dust particles
do not interact with each other and the action of
dust on the gas is neglected. \citet{2004ApJ...603..292G} found that small
particles move smoothly to the midplane,
while large bodies oscillate about the midplane with decreasing amplitude.

It should be noted that the grain sizes discussed here are larger than those
found in the  interstellar medium. This is consistent with the spectral
signature recently observed in the disk of CQ Tau \citep{2003A&A...403..323T}
suggesting the presence of large grains (a few centimetres), as well as  
evidence of dust processing and grain growth found in other disks
\citep[e.g.][]{Meeus2003,Apai2004}.

In this article, we present our new three dimensional gas $+$ dust code and
compare the results of disk simulations briefly to the work of
\citet{1977MNRAS.180...57W} and more extensively with
\citet{2004ApJ...603..292G} and \citet{2004ApJ...608.1050G}.
The code allows us to follow both the radial migration and vertical settling
of dust in high resolution, and the possibility of including additional 
physical processes such as turbulence, grain evolution, and radiative transfer
in a simple approximation or detailed equation of state. It has already been
applied in a preliminary study which uses scattered light as a diagnosis of
dust settling \citep{SF2A2004}.

In a separate project, we incorporate dust physics into a parallel N-body $+$
SPH code which calculates self-gravitational forces using a tree based data
structure \citep{MHM2003,HMMBG2005}. The data tree adds considerably to the
algorithmic complexity of the code, but means we can consider gravitational
stability problems.

In section 2, we present the basic equations for dust hydrodynamics and in
section 3 we describe our code.  In section 4 we describe our simulations:
the physical hypotheses, units and initial state, as well as the results.
Finally in section 5 we discuss our results in the light of the afore-mentioned
references.

\section{Dust dynamics}

A single particle in circular orbit of radius $r$ about a central body of
mass $M_\star$ will move with the Keplerian velocity
\mbox{$v_k = \sqrt{\mathrm{G}M_\star/r}$}, where $G$ is the gravitational
constant. Gas moving in a disk around the same central body will typically
orbit at a slower velocity due to the radial pressure gradient, which solid
bodies moving in the gas disk do not experience and so orbit
at a velocity different than that of the gas. The two phases interact
via a drag which slows down the dust and forces it to migrate
inwards to conserve angular momentum.

According to \citet{whipple1,whipple2} and 
\citet{1977MNRAS.180...57W}, small bodies are strongly coupled to the gas
and have about the same velocity field, whereas large bodies follow
marginally perturbed Keplerian orbits. Medium size particles experience
the strongest perturbation to their movement, with increased accretion to the
central object and vertical settling.

Before describing the equation of motion, we must explain the notations we
have adopted for densities. We use the subscripts d and g to denote dust and
gas respectively. We then define the intrinsic density
($\rho_\mathrm{g}$, $\rho_\mathrm{d}$) of a fluid and the void fraction
($\theta_\mathrm{g}$, $\theta_\mathrm{d}$) as the fraction of the volume
occupied by a given phase, such that $\theta_\mathrm{d}+\theta_\mathrm{g}=1$.
The density ($\hat{\rho}_\mathrm{g}$, $\hat{\rho}_\mathrm{d}$) of a phase in
a given volume of fluid is then given by
\begin{equation} \left\{\begin{array}{l} \hat{\rho}_\mathrm{g} = \theta_\mathrm{g}\,\rho_\mathrm{g}\\
\hat{\rho}_\mathrm{d} = \theta_\mathrm{d}\,\rho_\mathrm{d} .
\end{array}\right.
\end{equation}

Dust particles have a high intrinsic density (in our case
$\rho_\mathrm{d}=1$~g\,cm$^{-3}$), 
so a given mass of dust occupies a very small volume compared to the same
mass of gas. Therefore, gas fills almost all the volume whereas dust occupies
only a small portion of it, and:
\begin{equation}
\left\{\begin{array}{l}
\hat{\rho}_\mathrm{{g}} \approx \rho_\mathrm{{g}}\\
\hat{\rho}_\mathrm{{d}} \ll \rho_\mathrm{{d}}.
\end{array}\right.
\end{equation}

When the only force acting on the dust is drag, the equation of motion reads
\begin{equation}
\frac{\mathrm{d}\vec{v}_\mathrm{d}}{\mathrm{d}t}=
-\frac{\vec{v}_\mathrm{d}-\vec{v}_\mathrm{g}}{t_\mathrm{s}},\
\mathrm{with}\ t_\mathrm{s} = \frac{m v}{F_\mathrm{D}},
\label{dustdyn}
\end{equation}
where $\vec{v}_\mathrm{d}$ is the dust velocity, $\vec{v}_\mathrm{g}$ the
gas velocity, $t$ the time, $t_\mathrm{s}$ the stopping time and
$F_\mathrm{D}$ the drag force. The mass of the dust grain is
\begin{equation}
m=\frac{4 \pi}{3} \rho_\mathrm{d}\,s^3
\end{equation}
where we have assumed the grains are spherical with radius $s$. We use
$v=|\vec{v}_\mathrm{d}-\vec{v}_\mathrm{g}|$ to denote the velocity difference
between dust and gas phases at a particular point in space.  The stopping
time is the time it takes for a dust grain that starts with a velocity
$\vec{v}_\mathrm{d}$ to reach the gas velocity $\vec{v}_\mathrm{g}$.

The functional form of the drag force is determined by the Reynolds number,
$Re$, and by the ratio of the mean free path of gas molecules, $\lambda$,
to the radius of the grains, $s$
\citep[see][]{1977MNRAS.180...57W,1996A&A...309..301S}. The Reynolds number
is given by $Re=2s \rho_\mathrm{g} v/\eta$ where
$\eta=\rho_\mathrm{g} \lambda C_\mathrm{s} /2$ is the gas molecular viscosity
and $C_\mathrm{s}$ the sound speed. Thus $Re = 4 s v / \lambda C_\mathrm{s}$.

If $\lambda<4s/9$, we are in the Stokes regime and the drag is due to the
wake created by dust particles moving through the gas. The expression for the
drag force varies according to the Reynolds number:
\begin{equation}
F_\mathrm{D}=\left\{\begin{array}{lll}
24\,Re^{-1}\,F & \mathrm{for} & Re<1 \\
24\,Re^{-0.6}\,F & \mathrm{for} & 1<Re<800 \\
0.44\,F & \mathrm{for} & Re > 800 \\
\end{array}\right.
\end{equation}
with
\begin{equation}
F=\pi s^2 \rho_\mathrm{g} v^2/2.
\end{equation}

If $\lambda>4s/9$ and $Re<1$, we are in the Epstein regime and the drag
is due to thermal agitation and is given by:
\begin{equation}
F_\mathrm{D}=\frac{4 \pi}{3} \rho_\mathrm{g} s^2 v C_\mathrm{s}.
\end{equation}

\section{Description of the code}

We use the Smoothed Particles Hydrodynamics (SPH) algorithm, a Lagrangian
technique described by \citet{1992ARA&A..30..543M}. The SPH equations and
approximations have been rigorously established by \citet{Bicknell}.

Our code was based on that of \citet{1996MNRAS.279..402M}, originally
developed to study tidally unstable accretion disks in cataclysmic variables.
We have made several major changes: the configuration is for that of a
protoplanetary disk rather than a binary disk and we have modified
the algorithm for finding near neighbours so that the code can better
handle variable spatial resolution. 
Following the work of \citet{1995CoPhC..87..225M} and \citet{Maddison98},
we have added a second ``dust'' phase
in a full, self-consistent, fluid approach, contrary to other studies
using only test particles to describe the dust phase \citep[e.g.][]{Rice2004}.
We take into account the pressure exerted by gas on dust and by dust on gas,
as well as the drag force of gas on dust. 
In the results presented in this paper, we did not calculate the 
drag force of dust on gas because it is negligible in magnitude and
computationally time consuming,
but it is implemented into the code and can be turned on if required.
We do not review the SPH method as it has
been extensively described \citep[see, for example,][]{1992ARA&A..30..543M},
but stress the changes made to the code written by
\citet{1996MNRAS.279..402M} in the following subsections.

\subsection{Density}

As we are following the evolution of two fluids, we therefore have two
independent density equations. The
gas density is obtained by summation over only the gas neighbours
and the dust density is obtained through summation over the dust
neighbours. Using the subscripts $a$ and $b$ to refer to gas particles and
$i$ and $j$ for dust particles, we find:
\begin{equation}
\left\{\begin{array}{l@{=}l}
\hat{\rho}_a & \sum_b m_b W_{ab}\\
\hat{\rho}_i & \sum_j m_j W_{ij}
\end{array}\right. ,
\end{equation} 
where $W_{ij}=W(|\vec{r}_i-\vec{r}_j|)$ and $W$ is the cubic spline kernel 
commonly used in SPH.

\subsection{Smoothing length and link list}

The SPH smoothing length is usually given by
\begin{equation}
h=h_0\left(\frac{\rho_0}{\rho}\right)^{1/3},
\label{eq:firsth}
\end{equation}
to ensure a roughly constant number of neighbours (in our case gas plus dust
particles). Since our two types of particles have very different masses,
to avoid numerical error we therefore chose to use the number density
$n = \rho / m$ rather than the mass density $\rho$ in the calculation of $h$.
This approach was extensively studied by \citet{Ott2003} and found to be more
accurate.

In order to calculate the number density and subsequently the smoothing
length, we need
to find all the close neighbours of a given particle. As our code does not 
include self gravity, rather that using a tree code to find neighbours
\citep[see][]{1989ApJS...70..419H}, we use a less time consuming link list. 
In the original implementation of the code, the cells in the linked list
were $2h_\mathrm{max}$ in size, where $h_\mathrm{max}$ was the maximum value
of the smoothing length over the entire simulation \citep{1996MNRAS.279..402M}.
With such a cell size,
a given particle's neighbours laid either in the same link cell or in one
immediately neighbouring it. However, with $h$ varying by as much as
a factor 1\,000, the number of particles in some cells became very large
and the searches became inefficient.
We implemented an alternative link-list scheme due to Speith (private
communication) in which the size of the link cell is chosen to be
sufficiently small so that each cell only contains a few particles. This means
that, in order to find all the neighbours, the search has to be run over
a large number of cells. We nevertheless find this to be efficient compared
to the traditionally used link cell algorithm without introducing the
programming complexity of a tree structure.

\subsection{Drag Force}

For the gas, the equation of motion is given by:
\begin{equation}
\frac{\mathrm{d}\vec{v}_a}{\mathrm{d}t}=
\vec{P}_{ab}+\vec{M}_{aj}+\vec{D}_{aj}+\vec{G}_{a}.
\end{equation}
$\vec{P}_{ab}$ is the usual SPH internal pressure term except that 
each fluid is scaled by the volume it occupies, hence
\begin{equation}
\vec{P}_{ab}=-\sum_b m_b \left(\frac{P_a \theta_a}{\hat{\rho}_a^2} +
\frac{P_b \theta_b}{\hat{\rho}_b^2} + \Pi_{ab} \right) \vec{\nabla}_a W_{ab},
\end{equation}
where $\Pi_{ab}$ is the SPH artificial viscosity as described in
\citet{1996MNRAS.279..402M}. The volume scaling is
ensured by the multiplication by the void fraction $\theta$. 

The mixed pressure term,
\begin{equation}
\vec{M}_{aj}=-\sum_j m_j \frac{P_a \theta_j}{\hat{\rho}_a
\hat{\rho}_j }\vec{\nabla}_a W_{aj},
\end{equation}
is the pressure exerted on one fluid by the other.

The drag force term is given by:
\begin{equation}
\vec{D}_{aj}=\sigma\sum_j\frac{m_j K_{aj}}{\hat{\rho}_j\hat{\rho}_a}
\left(\frac{\vec{v}_{ja}\cdot\vec{r}_{ja}}{r_{ja}^2+\eta^2}\right)
\vec{r}_{ja} W_{ja},\
\mathrm{with}\
K_{aj}=\frac{\rho_a \theta_j c_a}{s},
\end{equation}
with $\sigma=1/D$, $D$ being the number of dimensions,
and $c_a$ the sound speed associated with particle $a$. The drag exerted
by dust on gas is negligible and time consuming so we have not included
it in these computations.

Finally, $\vec{G}_{a}$ is the gravity exerted by the central star only,
as we do not take self gravity into account. The derivation of the
$\vec{M}_{aj}$ and $\vec{D}_{aj}$ terms can be found in \citet{Maddison98}.

The dust equations are slightly different. The internal pressure
term disappears because the dust fluid is made of large bodies (compared to
gas molecules) that hardly ever encounter each other. These encounters are
better described by an SPH shock treatment than by an internal pressure. Thus
the dust equation of motion is given by:
\begin{equation}
\frac{\mathrm{d}\vec{v}_i}{\mathrm{d}t}=\vec{M}_{ib}+\vec{D}_{ib}+\vec{G}_{i},
\end{equation}
where
\begin{equation}
\vec{M}_{ib}=-\theta_i \sum_b m_b \frac{P_b }{\hat{\rho}_i\hat{\rho}_b }
\vec{\nabla}_i W_{ib}
\end{equation}
is the mixed pressure term exerted on dust,
\begin{equation}
\vec{D}_{ib} = - \sigma \sum_b \frac{m_b K_{ib}}{\hat{\rho}_b\hat{\rho}_i}
\left( \frac{\vec{v}_{ib}\cdot\vec{r}_{ib}}{r_{ib}^2+\eta^2}\right)
\vec{r}_{ib} W_{ib}
\end{equation}
is the drag force term exerted on dust, and $G_{i}$ is the gravity exerted
by the central star.

We use an implicit scheme which is iterative and hence a matrix inversion is
not necessary to derive the drag force. The equations are given by
\citet{Monaghan97} and can be found in Appendix~\ref{app:implicit}.
Note that we use the simpler implicit scheme of \citet{Monaghan97}, since
our tests of the Tischer algorithm were very slow and
did not give much better results.

\subsection{Timestep}

A timestep is defined for each physical force: the pressure
($\delta t_\mathrm{p}$), gravity ($\delta t_\mathrm{g}$), and drag
($\delta t_\mathrm{d}$) timesteps. The pressure timestep is defined so as to
verify the Courant condition and includes the viscosity timestep
\citep[see][]{Maddison98}:

\begin{equation}
\left\{\begin{array}{l}
\delta t_\mathrm{p}=\min_a\displaystyle\frac{h}{c_a+0.6\zeta\bar{c}_{ab}} \\
\delta t_\mathrm{g}=\min_a\displaystyle\sqrt{\frac{h}{|f_a|}}
\end{array}\right.,
\end{equation}
where subscripts $a$ and $b$ refer to two given SPH particles, $c_a$ is
the sound speed and $f_a$ the acceleration (i.e.\ the net force per unit mass)
of particle $a$, and $\zeta$ is the name we give to the SPH artificial
viscosity parameter usually called $\alpha$ to avoid any confusion
with the Shakura-Sunyaev $\alpha$ \citep{1973A&A....24..337S}.
While $\delta t_\mathrm{p}$ is much larger than $\delta t_\mathrm{g}$, 
the gravity subroutine is much less time consuming than the pressure 
subroutine and so we therefore use operator splitting.  This allows 
us to enter the gravity routine several times while only entering the 
pressure routine once.  The pressure defines the global timestep $\delta t$, 
such that $\delta t = 0.4 \delta t_\mathrm{p}$, while $\delta t_\mathrm{g}$ 
gives the number of times the gravity routine is run after a single run of 
the pressure routine, following \citet{1996MNRAS.279..402M}.

We also use operator splitting for the drag force because
$\delta t_\mathrm{d} \gg \delta t_\mathrm{g}$ but $\delta t_\mathrm{d}$ can get
much smaller than $\delta t_\mathrm{p}$ in case of strong drag (which happens
close to the central object where the gas density is highest and for the
smaller grain sizes). We choose ``by hand'' the number of times the drag
routine is run after a single run of the pressure routine:
$\delta t_\mathrm{d}$ decreases with the grain size and we find that, down to
1~mm, we just need to run it once and then 5 times for 100~$\mu$m, 25 times
for 10~$\mu$m and 100 times for 1~$\mu$m grains.

\subsection{Limitations}

The code treats turbulence in a very limited manner.
The dissipation term in the equation of motion was chosen to reproduce the 
level of dissipation characteristic of the Shakura-Sunyaev turbulence model.
However, the length scale on which dissipation occurs is of the order of the 
smoothing length and so a full turbulent cascade does not develop.

The resolution of the inner edge of the disk is also limited.
In the calculations, the disk scale height, which is defined by
$H=C_\mathrm{s}/\Omega$, where $\Omega=\sqrt{GM_\star/r^3}$ is the angular
velocity,
has to be larger than the smoothing length $h$ otherwise the disk is not
resolved. However, $H<h$ at small radii and we therefore must set
$H_\mathrm{code}=\max(H,h)$ to be able to carry on the calculations. 
It should be noted therefore that the disk is unresolved 
for $r < 6$~AU in our high resolution simulations and 
$r< 20-40$~AU in the low resolution simulations.
Indeed, our disk extends over more than 2~orders of magnitude in radius 
making it difficult to resolve both the small and large radii.
The solution is not, however, to simply truncate the disk at 6~AU (in the 
high resolution case) because this would result in boundary layer problems.
Instead we know that we can trust the results from 6~AU.
Hereafter, what we call the inner parts of the disk refer to the regions
immediately outside the unresolved parts and not the actual disk inner edge.

\section{Simulations}

We model a protoplanetary disk of mass 0.01~$M_\odot$ with 1\% of dust by mass
around a 1~$M_\odot$ star. We consider dust particles ranging from 1~$\mu$m
to 10~m and run a series of simulations that include only one grain size at a
time. Our disk is vertically isothermal which means that the temperature is
constant in the vertical direction but follows a radial power law 
($T \propto r^{-3/4}$). We implicitly assume that whatever heating process is
acting within the disk, the subsequent heat is immediately radiated away.

We do not include the effects of photoevaporation, radiation pressure,
Poynting Robertson flux, magnetic fields or grain coagulation, sublimation
or condensation.

We scale our model so as to have numbers close to unity. The mass unit is
one solar mass, the length unit 100~AU and the
time unit is the orbital period of a particle at 100~AU around a 1~$M_\odot$
object divided by $2\pi$ (1000~yr/$2\pi$). This choice of units leads
to a gravitational constant set to 1.

We choose an initial state in near equilibrium conditions. In the
following simulations, we let a gaseous non dusty disk relax from an
initial distribution given by $\Sigma=\Sigma_0 r^{-3/2}$ and
$H=C_\mathrm{s}/\Omega$. The initial velocity is Keplerian, given by
$v_k=\sqrt{GM_\star/r}$.
For the power laws of the temperature and initial surface density, we take the
parameters of the Minimum Mass Solar Nebula \citep{Hayashi85}.

Starting from that initial distribution, we let the disk relax for
$\sim$ 8\,000~years, i.e. 8~orbits at 100~AU, which allows the pressure and
artificial viscosity time to smooth out the velocity field. The gas velocity
becomes slightly sub-Keplerian because of the pressure gradient. Once the gas
disk has relaxed, we then add the dust particles on top of the gas particles
with the same velocity and let the disk evolve for a further 8\,000 years.

\begin{figure*}
\centering
\includegraphics[width=15cm]{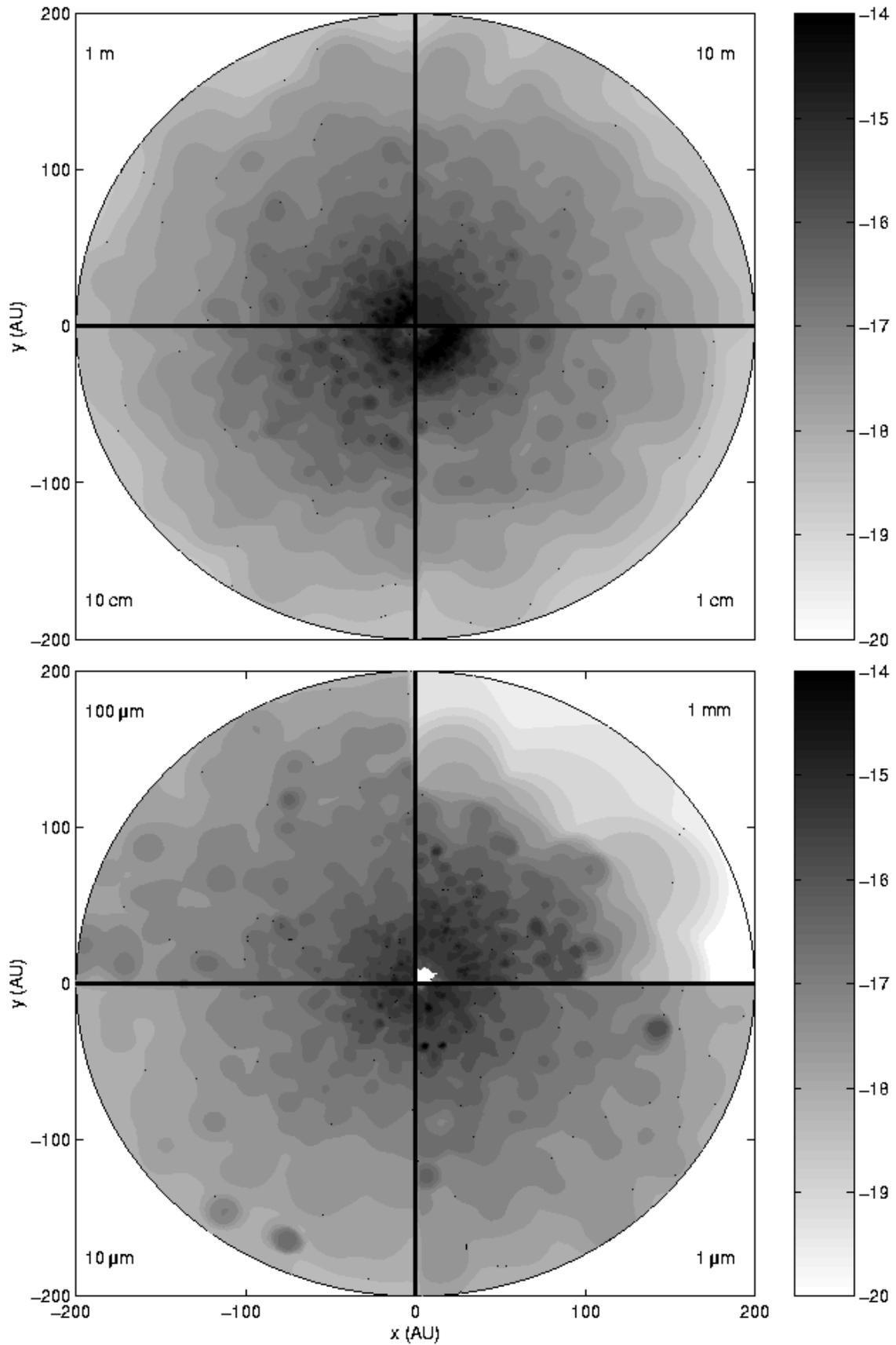}
\caption{Density contours for dust particles for the disk seen face on
at the end of the low resolution simulations, each quadrant representing a
different size of particles. The vertical bar gives $\log\rho$ in
g\,cm$^{-3}$.}
\label{face}
\end{figure*}

\begin{figure*}
\sidecaption
\includegraphics[width=12cm]{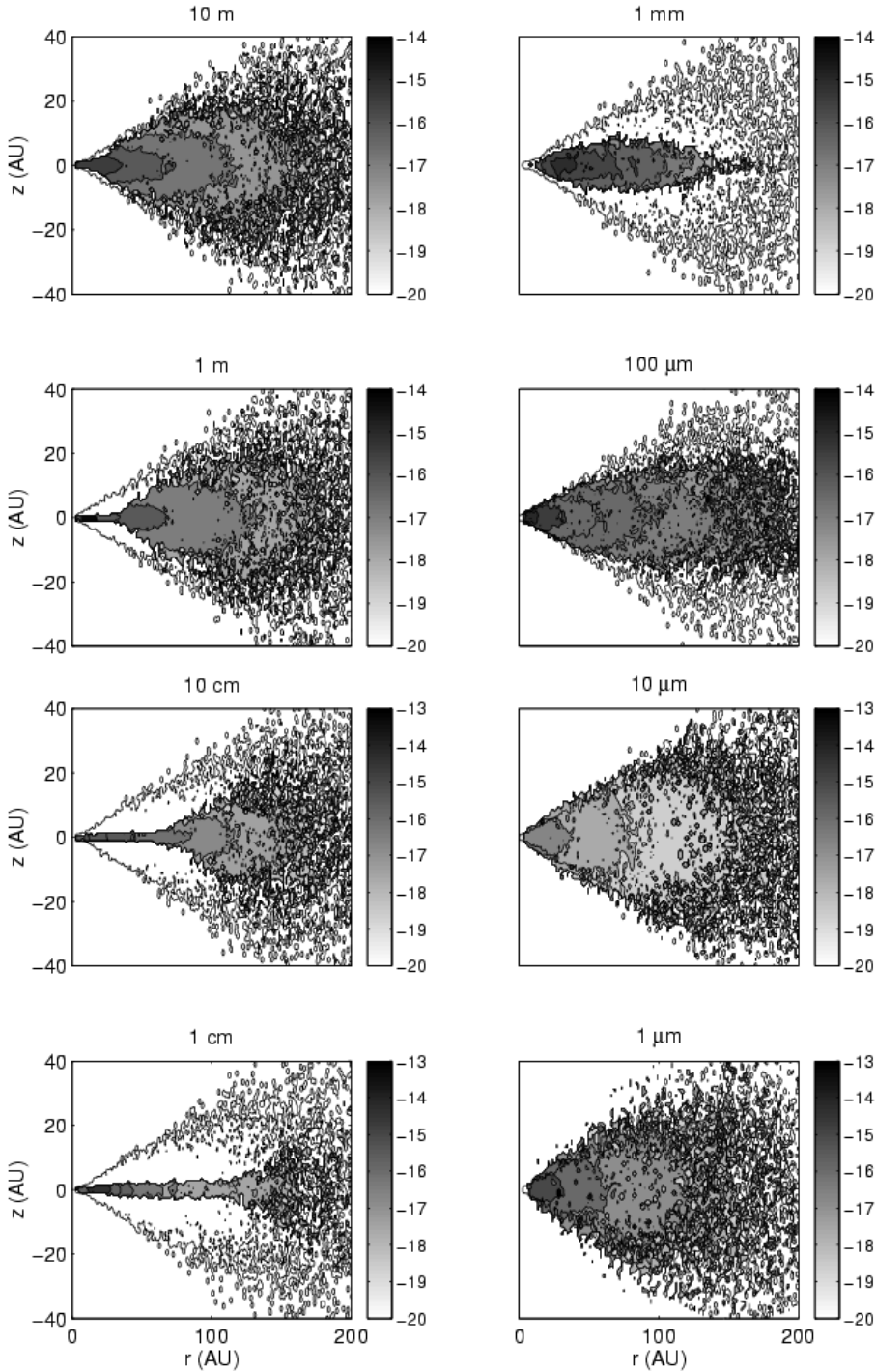}
\caption{Density contours for dust particles of different sizes for the disk
seen edge on at the end of the low resolution simulations. The outermost
contour is the $\rho=1.9\,10^{-22}$~g\,cm$^{-3}$ contour for the gas. The
vertical bar gives $\log\rho$ in g\,cm$^{-3}$.}
\label{profil}
\end{figure*}

We checked that the Reynolds number is always less than
unity since the velocity difference between dust and gas particles stays
subsonic. The mean free path is given by
\begin{equation}
\lambda=\frac{1}{n \sigma_\mathrm{0}} \approx \frac{m_\mathrm{H_2}}
{\pi \rho_\mathrm{g} r^2_\mathrm{H_2}}.
\end{equation}
The right hand side is obtained using the approximations made by
\citet{1996A&A...309..301S}, that hydrogen is considered as molecular and the
cross section is assumed to be identical to the geometrical section.
$\lambda$ decreases with increasing density, so if the criterion for the
Epstein regime, $\lambda>4s/9$, is met in the densest regions
(i.e. the centre of the disk) then it will be met everywhere.
Because our disk is not massive ($M_\mathrm{disk}=0.01 \mathrm{M_\odot}$)
and is radially extended (300~AU), the highest gas volumetric density we get
is of about $5\,10^{-11}$~g\,cm$^{-3}$, giving $9 \lambda / 4 > 11.5$~m.
Therefore, for $10^{-6}~\mathrm{m} < s < 10$~m, we are in the Epstein regime,
and
\begin{equation}
t_\mathrm{s}=\frac{s \rho_\mathrm{d}}{C_\mathrm{s} \rho_\mathrm{g}}
\end{equation}
and the equation of motion for the dust is given by
\begin{equation}
\frac{\mathrm{d}\vec{v}_\mathrm{d}}{\mathrm{d}t}=-\frac{C_\mathrm{s}
\rho_\mathrm{g}}{s \rho_\mathrm{d}}(\vec{v}_\mathrm{d}-\vec{v}_\mathrm{g}).
\end{equation}

We studied a large range of dust grains sizes at low resolution
($\sim 25\,000$ SPH particles) to see the effect of grain size on the dust
settling morphology. The simulations proved to have evolved for long enough 
to see the most striking features of
the dust settling. We then focused on intermediate size grains (10~cm to
100~$\mu$m) where the drag is most efficient and ran high resolution
simulations ($\sim 160\,000$ SPH particles). The results are shown on
Figs.~\ref{face} and \ref{profil} for the low resolution, and
Table~\ref{param} lists the simulation parameters.

\begin{table}
\caption{Simulation parameters}
\label{param}
\begin{tabular}{@{}l@{\ =\ }ll@{\ =\ }l@{}}
\hline\hline
$M_\star$ & $1.0\ M_\odot$ &
$s$ & 10,1,10$^{-1}$,10$^{-2}$,10$^{-3}$,10$^{-4}$,10$^{-5}$,10$^{-6}$ m \\
$M_\mathrm{gas}$ & $0.01\ M_\star$ &
$C_\mathrm{s}$ & $C_0\,r^{-3/8}$, $C_0=0.1$ code units \\
$M_\mathrm{dust}$ &$0.01\ M_\mathrm{gas}$ & $\Sigma$ & $\Sigma_0\,r^{-3/2}$ \\
$R_\mathrm{disk}$ & 300 AU & $\alpha_\mathrm{SPH}$ & $\zeta=0.1$ \\
$\rho_\mathrm{d}$ & 1.0 g\,cm$^{-3}$ & $\beta_\mathrm{SPH}$ & 0.0 \\
\hline
\end{tabular}
\end{table}

A high resolution single grain size computation requires an average of 
15\,000~CPU hours shared over 64~processors with OpenMP parallelisation
on the SGI Origin 3\,800 of the French national computing centre CINES
based in Montpellier. As the grain size decreases, the drag timestep gets
smaller and the computation time increases. The 100~$\mu$m simulation required
$22\,400$ CPU hours.

\section{Discussion}
\label{Discussion}

Our simulations show that the dust in the solar nebula behaves qualitatively
in the way of the analysis of \citet{whipple1,whipple2} and
\citet{1977MNRAS.180...57W}. For example, the larger bodies ($s \ge 1$~m) are
weakly
coupled to the gas and follow marginally perturbed Keplerian orbits and the
dust disk is flared except near the centre where the gas drag is the most
efficient because of the high gas density.  On the other hand, the smaller
bodies ($s \le 10\ \mu$m) are so strongly coupled to the gas that they follow
its motion and again the dust disk is flared.

\begin{figure}
\resizebox{\hsize}{!}{\includegraphics[angle=-90]{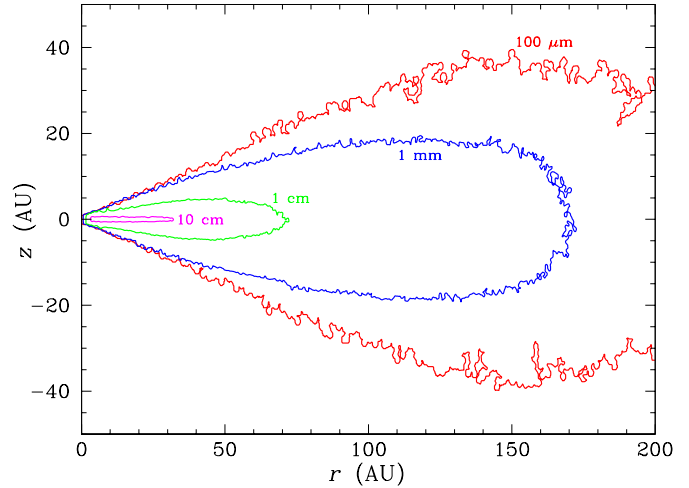}}
\caption{$\mu=1$ contours (see text) at the end of the high-resolution
simulations for 100~$\mu$m, 1~mm and 1~cm grains.
For the 10~cm particles, this value was not reached and the contour is
drawn for $\mu=0.5$ instead.
}
\label{mu}
\end{figure}

But for intermediate size particles (100~$\mu$m $\le s \le$ 10~cm), the drag
strongly influences the dust
dynamics and induces a motion that is very different from that of the gas. We
first see a rapid stopping phase, where the dust settles into a region where
the drag force dominates. This region is shown in Fig.~\ref{mu} as the
interior of the $\mu=1$ contours. This parameter $\mu$ was characterised
by \citet{2004ApJ...603..292G} as the ratio of the orbital time to the
stopping time:
\begin{equation}
\mu=\frac{1}{\Omega}\frac{\rho_\mathrm{g}}{\rho_\mathrm{d}}\frac{C_\mathrm{s}}{s}.
\label{eq:mu}
\end{equation}
When $\mu \gg 1$, the Epstein drag is very strong. Once the larger grains
reach the $\mu\ge1$ region, they experience rapid settling to the midplane as
well as strong radial accretion and fall in the central star. The dust
layer therefore gets very thin at small radii. For the smaller grains,
the settling to the midplane is so efficient that radial migration cannot
respond fast enough and there is a particle pileup and the
dust layer gets thicker again. 
Unfortunately, we cannot quantitatively predict
the thickness of the dust layer with these simulations, as dispersion in
the velocities of individual SPH particles limits our resolution.

Figure~\ref{face} also shows the transition between regimes of weak and
strong vertical settling, as well as weak and strong radial migration. The 10~m
particles do not show any particular increase in density because the
settling is weak, and their distribution stays very close to their initial
configuration, which is that of the gas (as one can see in the edge-on plots in
Fig.~\ref{profil}). Then for 1~m particles, the settling becomes efficient
in the inner regions of the disk, but the radial migration is still weak.
The particles therefore pile up towards the inner edge of the disk and the
density increases. For 10~cm particles, the settling is even stronger and
extends to larger radii, but then the radial migration becomes efficient and
the particles are lost to the star and hence the density enhancement is
weaker than for larger grains. For 1~cm
particles, the settling is now so efficient that it extends over even larger
radii and despite efficient accretion, too many particles arrive at the
inner part of the disk at the same time and thus they pile up and we see a
strong density enhancement. For 1~mm
particles, the combined effects of settling and migration have proven so
efficient that the outer parts of the disk are depleted of dust.
Then the efficiency of the vertical settling decreases together with the
radial migration, up to the point where particles are completely
coupled to the gas as for 1~$\mu$m particles, for which the distribution is
undistinguishable from their inital configuration.

\begin{figure}
\resizebox{\hsize}{!}{\includegraphics{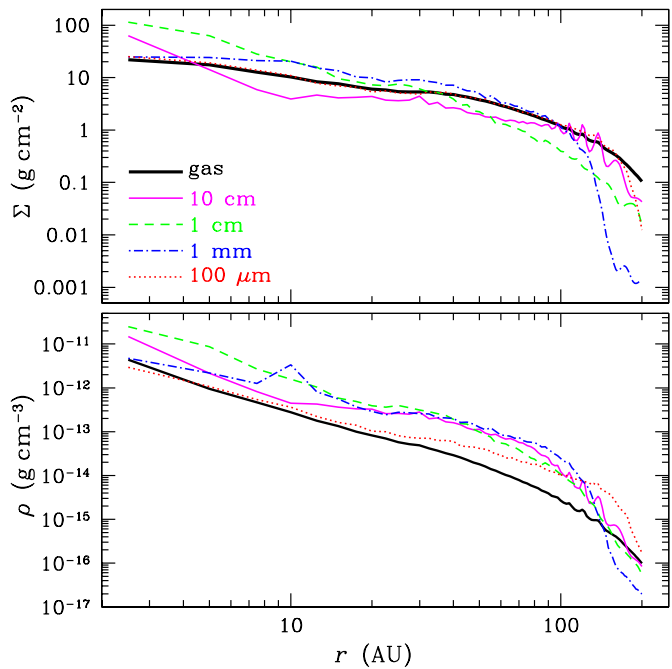}}
\caption{Surface density $\Sigma$ (top panel) and volume density $\rho$
(bottom panel) as a function of $r$ at the end of the high-resolution
simulations. In each case, the dust density is multiplied by 100 to better
view the dust density increase.}
\label{densities}
\end{figure}

In Fig.~\ref{densities}, one can see a large increase in the volume density
ratio between dust and gas, rising from $10^{-2}$ to
$10^{-1}$ for some radii. This increase is not as striking when we
look at the surface density because the effect of the settling is hidden by
the summation over the vertical scale height of the disk. Therefore
2D simulations will underestimate the density enhancement. The meaningful
quantity to study is in fact the volumetric density, which has the potential 
to sufficiently damp the dust velocities to allow the grains to 
stick together via collisions instead of being shattered. 
It should be noted, however, that while we do not consider the drag of the 
dust on the gas because it is negligible when the dust to gas ratio is small, 
as this ratio approach one the dust could drag the gas with it
into high density regions, thus resulting in a self regulation of the dust
to gas density ratio.

\citet{HMMBG2005} show that, for values typical of the solar nebula, the
distribution of solid particles evolves to an essentially stationary state
and that the final radius of the dust disk can be characterised 
in terms of the particle size.

As in \citet{1977MNRAS.180...57W}, we find that the radial migration rate is
highest for a given grain size and drops quickly for larger and smaller
particle sizes. We find this maximum is reached for 1~cm size particles,
whereas \citet{1977MNRAS.180...57W} found it to be 1~m. This discrepancy comes
from the difference in the parameters we use for the nebula and
\citet{1977MNRAS.180...57W} showed that the gas density and dust intrinsic
density have an impact on this maximum grain size.

\begin{figure}
\resizebox{\hsize}{!}{\includegraphics{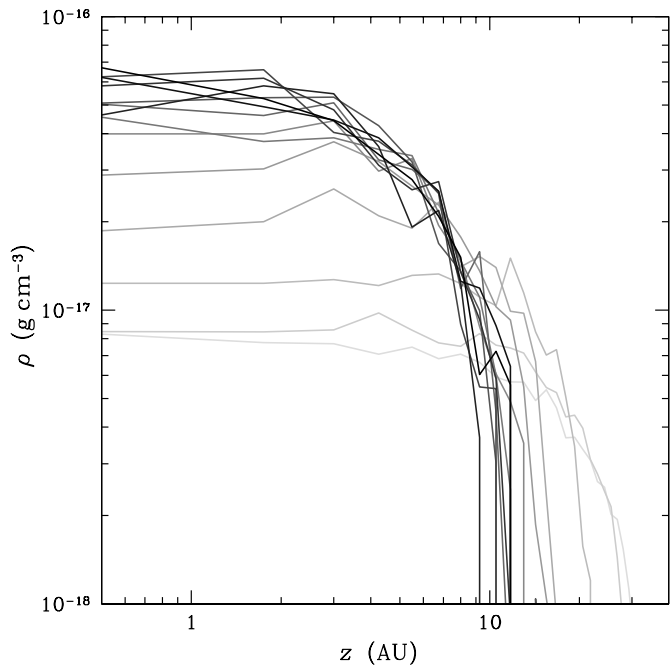}}
\caption{Volume density versus $z$ at $r=150$~AU for the 100~$\mu$m dust at
different times from dust injection (lightest curve) until the end of the
high-resolution simulations (darkest curve).}
\label{rho_z}
\end{figure}

To check the vertical evolution, we first compare our results with the
self-similar behaviour of the dust phase during settling as described by
\citet{2004ApJ...608.1050G} and shown in their figures~1b and 2. Despite the
noise due to the SPH technique, our results indeed show the same kind of
behaviour, shown in Fig.~\ref{rho_z}. Note that our density increase is
orders of magnitudes smaller than theirs because the computation is too time
consuming to be run for as long an evolutionary time as given in
\citet{2004ApJ...608.1050G}.

\begin{figure*}
\sidecaption
\includegraphics[width=12cm]{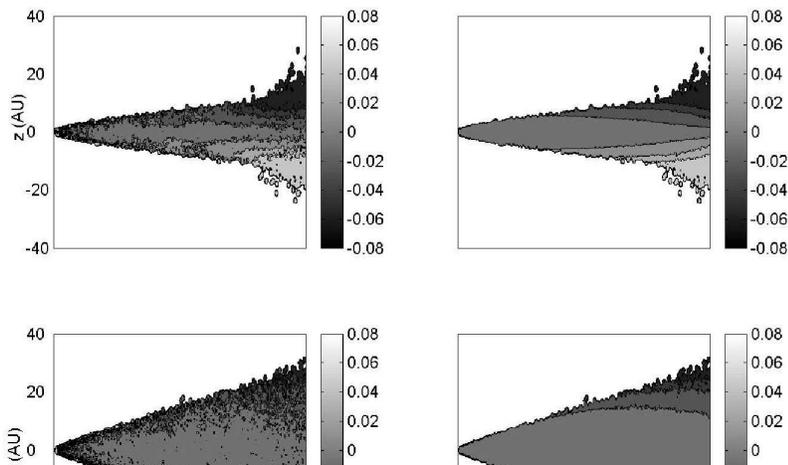}
\caption{Contours of the vertical velocity for 1~mm (top panels) and 100~$\mu$m
(bottom panels) particles for the high-resolution simulations 400~yrs after
dust injection. Left: data from our simulations, right: expected theoretical
values from Eq.~(\ref{eq:vz}). The vertical bar gives $v_z$ in code units
(see text).}
\label{cont_vz}
\end{figure*}

We next validate the formula derived by \citet{2004ApJ...603..292G} in the
Epstein regime and reproduce their figure~5 with our conditions
(see Fig.~\ref{cont_vz}). Starting from the trajectory of a single
particle and then using a Boltzmann averaging, \citet{2004ApJ...603..292G}
derived an analytical formula for the vertical velocity
\begin{equation}
v_z=-\frac{z}{\mu},
\label{eq:vz}
\end{equation}
with $\mu$ defined in Eq.~(\ref{eq:mu}). To calculate the value of $\mu$
and then of $v_z$, we use the gas density that comes out of the simulations
and the sound speed prescription that we use in our code. We then
compare this computed value of $v_z$ with the one taken directly from the
simulations in Fig.~\ref{cont_vz}.

The proportionality between vertical velocity and vertical position is
most striking at the very beginning of the simulation when the vertical
extension is maximal and the particle velocities towards the midplane are
largest. The plots are shown 400~yrs (0.4~orbits at 100~AU) after the addition
of dust.

We can easily see the vertical stratification of the velocity for the 1~mm
grains. It is less obvious for the 100~$\mu$m grains because the dust 
is more strongly coupled to the gas and settles more slowly towards the
midplane. Furthermore, the noise from the simulation makes the structure more
difficult to detect. At the beginning of the simulations for the 10~cm and
1~cm particles, the grains oscillate about the midplane
following perturbed Keplerian orbits before
the gas damps these oscillations efficiently as shown on figure~2 of
\citet{2004ApJ...603..292G}.

Finally, looking at the sequence of edge-on images in Fig.~\ref{profil}, we
see qualitatively the behaviour discussed by \citet{2004ApJ...603..292G} 
that at a given height the region will be depleted of
intermediate size particles.

\section{Conclusion}

\subsection{Summary}

In this article, we have shown the importance of the vertical settling of
intermediate size grains (10~cm to 100~$\mu$m) in a protoplanetary disk
with a central object of $M_\star$ = $1.0~M_\odot$, a gas disk mass of
$M_\mathrm{gas}=0.01~M_\star$ and a dust disk mass of
$M_\mathrm{dust}=0.01~M_\mathrm{gas}$.

We find that the dust distribution is completely different from the gas and the
general approximation that the dust to gas ratio is constant throughout
the disk is wrong for grains in this size range. This has an impact on the
interpretation of observations made in the millimetre domain, for instance, 
as the opacity of millimetre size grains dominates. It is thus important to
understand the specific behaviour of the dust to avoid misinterpretations such
as an underestimate or overestimate of the gas density.

Our simulation results compare well with the radial migration seen in
\citet{1977MNRAS.180...57W}, the
analytical work of \citet{2004ApJ...603..292G}
and the vertical settling of \citet{2004ApJ...608.1050G}.

\subsection{Perspectives}

The \citet{1973A&A....24..337S} turbulence modelling has been extensively
used in the domain of protoplanetary disks and has proven to give
consistent results. It should, however, be used cautiously and ideally 
a better prescription for turbulence should be used.
Indeed, turbulence will likely reduce the dust settling velocities and lessen
the density enhancements we observe in our simulations
\citep[see, e.g.,][]{1993prpl.conf.1031W}.
The magneto-rotational instability has been consistently described by
\citet{1991ApJ...376..214B} and global numerical simulations for this
instability are becoming more accurate \citep{2004ApJ...616..364F}.
It is now possible to have a consistent description of this kind of turbulence
in a numerical simulation and a  first step towards this goal was given by
\citet{2002MNRAS.335..843M}, who  derived an SPH prescription of turbulence.
We plan to incorpate SPH turbulence into our code 
in order to investigate how turbulence affects the dust settling.

While the locally isothermal approximation is roughly consistent with the
temperature distribution in a protoplanetary disk, a lot of effort is
currently going into the domain of radiative transfer in protoplanetary disks
\citep[e.g.][]{2004A&A...421.1075D}, suggesting
that the energetics in such an object deserve a better treatment.
Unfortunately, including full radiative transfer in our SPH code is 
not feasible because  of the increased computation it would require.
Nevertheless,
we will be able to achieve a better description by implementing an adiabatic
equation of state with cooling functions as one of the improvements we plan
to add in our code.

This consistent description of the dust distribution in a protoplanetary disk
is a first step towards planet formation. By implementing a consistent
treatment of the grain growth, coagulation, and shattering, we will be able to
better understand planetesimal formation and their distribution in the
disk.

\begin{acknowledgements}
This work was funded by the French PNPS and L. B.-F. would like to thank
the French CINES for the important amount of computing time that made these
simulations possible. The authors are grateful to the French-Australian PICS
for financial support for travel to Australia. L. B.-F. is also deeply
grateful to Eduardo Delgado for comments and suggestions as well as to
Roland Speith for the idea of the new link list routine.
\end{acknowledgements}

\bibliographystyle{aa}
\bibliography{2249bib}

\appendix

\section{Implicit schemes for the drag force}
\label{app:implicit}

All the following equations are derived from \citet{Monaghan97}.
The momentum equations for gas and dust in Epstein regime read
\begin{equation}
\frac{\mathrm{d}\vec{v}_\mathrm{g}}{\mathrm{d}t}=
-\frac{K}{\hat{\rho}_\mathrm{g}} (\vec{v}_\mathrm{g}-\vec{v}_\mathrm{d})
\end{equation}
\begin{equation}
\frac{\mathrm{d}\vec{v}_\mathrm{d}}{\mathrm{d}t}=
-\frac{K}{\hat{\rho}_\mathrm{d}} (\vec{v}_\mathrm{d}-\vec{v}_\mathrm{g}),
\end{equation}
with
\begin{equation}
K=\frac{C_\mathrm{s} \hat{\rho}_\mathrm{g} \hat{\rho}_\mathrm{d}}
{\rho_\mathrm{d} s},
\end{equation}
and become in SPH notations (indices $a$ and $b$ refer to gas particles
and $i$ and $j$ to dust particles and for a vector $\vec{q}$,
$\vec{q}_{aj} = \vec{q}_a - \vec{q}_j$):
\begin{equation}
\frac{\mathrm{d}\vec{v}_a}{\mathrm{d}t}=\sigma \sum_j m_j
\frac{K_{aj}}{\hat{\rho}_j \hat{\rho}_a}\left(\frac{\vec{v}_{aj}\cdot
\vec{r}_{aj}}{r_{ja}^2 + \eta^2} \right) \vec{r}_{aj} W_{aj}
\end{equation}
\begin{equation}
\frac{\mathrm{d}\vec{v}_j}{\mathrm{d}t}=\sigma \sum_a m_a
\frac{K_{aj}}{\hat{\rho}_a \hat{\rho}_j }\left(\frac{\vec{v}_{aj}\cdot
\vec{r}_{aj}}{r_{aj}^2 + \eta^2} \right) \vec{r}_{ja} W_{ja}.
\end{equation}
Then, it is possible to turn it to an implicit scheme without having to
invert a matrix. We have
\begin{equation}
\frac{\mathrm{d}\vec{v}_a}{\mathrm{d}t}=
\sum_j m_j s_{aj} (\vec{v}_{aj} \cdot \vec{r}_{aj}) \vec{r}_{aj} 
\end{equation}
\begin{equation}
\frac{\mathrm{d}\vec{v}_j}{\mathrm{d}t}=
\sum_a m_a s_{aj} (\vec{v}_{aj} \cdot \vec{r}_{aj}) \vec{r}_{ja},
\end{equation}
with
\begin{equation}
s_{aj}=\sigma \frac{K_{aj}}{\hat{\rho}_a \hat{\rho}_j}
\left(\frac{W_{aj}}{r_{aj}^2 + \eta^2} \right).
\end{equation}
With a timestep $\delta t$ and $\vec{v}_i^0$ and $\vec{v}_i^1$ respectively
the old and new values of the velocity:
\begin{equation}
\vec{v}_a^1 = \vec{v}_a^0 - \delta t \sum_j m_j s_{aj}
(\vec{v}_{aj}^1 \cdot \vec{r}_{aj})\vec{r}_{aj}
\end{equation}
\begin{equation}
\vec{v}_j^1 = \vec{v}_j^0 - \delta t \sum_a m_a s_{aj}
(\vec{v}_{aj}^1 \cdot \vec{r}_{aj})\vec{r}_{ja}.
\end{equation}
Now, we will just consider one dust-gas particle pair at a time:
\begin{equation}
\vec{v}_a^1 = \vec{v}_a^0 - \delta t\,m_j s_{aj}
(\vec{v}_{aj}^1 \cdot \vec{r}_{aj})\vec{r}_{aj}
\end{equation}
\begin{equation}
\vec{v}_j^1 = \vec{v}_j^0 - \delta t\,m_a s_{aj}
(\vec{v}_{aj}^1 \cdot \vec{r}_{aj})\vec{r}_{ja}.
\end{equation}
Then taking the difference of the previous two equations and taking the scalar 
product with $\vec{r}_{aj}$, we get
\begin{equation}
\vec{v}_{aj}^1 \cdot \vec{r}_{aj} = \frac{\vec{v}_{aj}^0 \cdot \vec{r}_{aj}}
{1 + \delta t  (m_a + m_j) s_{aj} r_{ja}^2},
\end{equation}
and
\begin{equation}
\vec{v}_a^1 = \vec{v}_a^0 - \frac{\delta t m_j s_{aj}\vec{r}_{aj}
(\vec{v}_{aj}^0 \cdot \vec{r}_{aj})}{1+\delta t (m_a + m_j) s_{aj}r_{ja}^2}
\end{equation}
\begin{equation}
\vec{v}_j^1 = \vec{v}_j^0 - \frac{\delta t m_a s_{aj}\vec{r}_{ja}
(\vec{v}_{aj}^0 \cdot \vec{r}_{aj})}{1+\delta t (m_a + m_j) s_{aj}r_{ja}^2}.
\end{equation}

The drag of dust on gas is very small, and to save computing time, we
neglect it. So, at the end, only dust experiences drag with the following
 expression:
\begin{equation}
\vec{v}_j^1 = \vec{v}_j^0 - \sum_j \frac{\delta t m_a s_{aj}\vec{r}_{ja}
(\vec{v}_{aj}^0 \cdot \vec{r}_{aj})}{1+\delta t (m_a + m_j) s_{aj}r_{ja}^2}.
\end{equation}

\subsection{Tischer implicit scheme}

Now, we subdivide each timestep into two, and we note $\vec{\tilde{v}}$
the velocity at $\Delta = \delta t / 2$:
\begin{equation}
\vec{\tilde{v}}_a = \vec{v}_a^0 - \Delta\vec{r}_{aj} m_j s_{aj}
(0.6\vec{\tilde{v}}_{aj}\cdot\vec{r}_{aj}+0.4\vec{v}_{aj}^0\cdot\vec{r}_{aj}),
\end{equation}
\begin{equation}
\vec{\tilde{v}}_j = \vec{v}_j^0 - \Delta\vec{r}_{aj} m_j s_{aj}
(0.6\vec{\tilde{v}}_{aj}\cdot\vec{r}_{aj}+0.4\vec{v}_{aj}^0\cdot\vec{r}_{aj}),
\end{equation}
and 
\begin{equation}
\vec{v}_a^1 = 1.4\vec{\tilde{v}}_a - 0.4\vec{v}_a^0 - 0.6\Delta\vec{r}_{aj}
m_j s_{aj}(\vec{v}_{aj}^1 \cdot \vec{r}_{aj}),
\end{equation}
\begin{equation}
\vec{v}_j^1 = 1.4\vec{\tilde{v}}_j - 0.4\vec{v}_j^0 - 0.6\Delta\vec{r}_{aj}
m_a s_{aj}(\vec{v}_{aj}^1 \cdot \vec{r}_{aj}).
\end{equation}
For simpler notations, we define $A=\Delta r^2_{aj} (m_j+m_a) s_{aj}$ and
$B=\Delta r_{aj} m_a s_{aj}$. Then
\begin{equation}
\vec{\tilde{v}}_{aj}\cdot\vec{r}_{aj}=
\frac{1-0.4A}{1+0.6A}\vec{v}^0_{aj}\cdot\vec{r}_{aj}, 
\end{equation}
\begin{equation}
\vec{v}^1_{aj}\cdot\vec{r}_{aj}=
\frac{1-0.8A}{(1+0.6A)^2}\vec{v}^0_{aj}\cdot\vec{r}_{aj},
\end{equation}
and
\begin{equation}
\vec{v}^1_{j}=\vec{v}^0_{j}+B\vec{r}_{aj}\frac{2+0.36A}{(1+0.6A)^2}
\vec{v}^0_{aj}\cdot\vec{r}_{aj}.
\end{equation}

\end{document}